\documentclass{article}

\usepackage{microtype}
\usepackage{graphicx}
\usepackage{subfigure}
\usepackage{booktabs} 
\usepackage{physics}

\usepackage{hyperref}

\newcommand{\psixt}{\psi(\mathbf{x},t)}

\usepackage[accepted]{icml2021}

\icmltitlerunning{Spacetime Neural Network for High Dimensional Quantum Dynamics}

\begin{document}

\twocolumn[
\icmltitle{Spacetime Neural Network for High Dimensional Quantum Dynamics}

\icmlsetsymbol{equal}{*}

\begin{icmlauthorlist}
\icmlauthor{Jiangran Wang}{ece}
\icmlauthor{Zhuo Chen}{phys}
\icmlauthor{Di Luo}{phys,icmt}
\icmlauthor{Zhizhen Zhao}{ece}
\icmlauthor{Vera Mikyoung Hur}{math}
\icmlauthor{Bryan K. Clark}{phys,icmt}
\end{icmlauthorlist}

\icmlaffiliation{ece}{Department of Electrical and Computer Engineering and CSL, University of Illinois at Urbana-Champaign, Urbana, IL 61801, USA}
\icmlaffiliation{phys}{Department of Physics, University of Illinois at Urbana-Champaign, IL 61801, USA}
\icmlaffiliation{icmt}{IQUIST and Institute for Condensed Matter Theory and NCSA Center for Artificial Intelligence Innovation,  University of Illinois at Urbana-Champaign, IL 61801, USA}
\icmlaffiliation{math}{Department of Mathematics, University of Illinois at Urbana-Champaign, Urbana, IL 61801, USA}

\icmlcorrespondingauthor{Di Luo}{diluo2@illinois.edu}
\icmlkeywords{Machine Learning, ICML}

\vskip 0.3in
 ]

\printAffiliationsAndNotice{}  

\begin{abstract}
We develop a spacetime neural network method with second order optimization for solving quantum dynamics from the high dimensional Schrödinger equation. In contrast to the standard iterative first order optimization and the time-dependent variational principle, our approach utilizes the implicit mid-point method and generates the solution for all spatial and temporal values simultaneously after optimization. We demonstrate the method in the Schrödinger equation with a self-normalized autoregressive spacetime neural network construction. Future explorations for solving different high dimensional differential equations are discussed.
\end{abstract}

\section{Introduction}

\begin{figure*}[h!]
\centering
\includegraphics[width=1.0\linewidth]{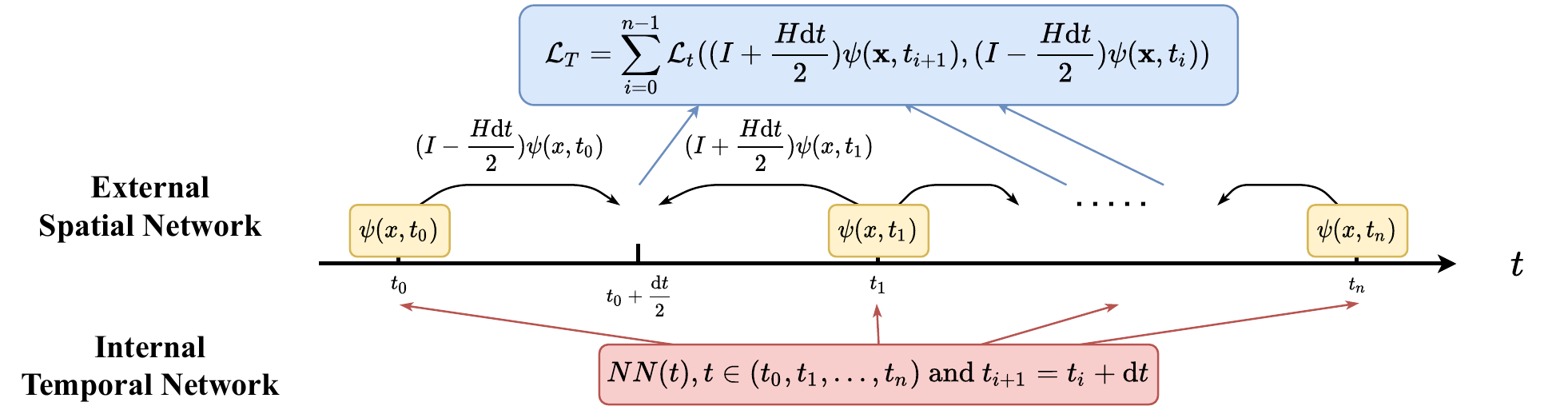}
\caption{Second order optimization with spacetime neural network. The spacetime neural network consists of an internal temporal network and an external spatial network. The internal temporal network provides time dependent weights for the external spatial network at each time step. The loss function is computed over all time steps and is used to train the spacetime neural network.}
\label{fig:spacetime}
\end{figure*}

Differential equation plays a fundamental role in science and engineering. Among different differential equations, the Schrödinger equation is a famous example of high dimensional differential equations that describes quantum dynamics of a physical system. The real time Schrödinger equation is given by
\begin{equation}
    i\frac{\partial \psi(\mathbf{x}, t)}{\partial t} = H \psi(\mathbf{x}, t)
\end{equation} 
and the imaginary time version is 
\begin{equation}
    \frac{\partial \psi(\mathbf{x}, t)}{\partial t} = -H \psi(\mathbf{x}, t)
\end{equation} 
where $\psi(\mathbf{x},t)$ is a complex-valued function whose dimensionality grows exponentially with respect to the number of particles ($\mathbf{x}$ is the configuration of the particles) and $H$ is the Hamiltonian of the physical system.  

With the recent advancement of quantum science and engineering, the research of quantum dynamics is becoming an important topic. Quantum dynamics takes place in various areas, such as photosynthesis, chemical reactions, cold atom experiments, and quantum computation~\cite{Romero2014,Clary1879, Cirac2012,Arute2019}. New approaches for solving the Schr{\"o}dinger equation will provide powerful tools to understand and explore quantum dynamics with various applications. 

In recent years, the advancement of machine learning opens up new possibilities for solving the high dimensional Schrödinger equation~\cite{Carleo_2017}.  The standard approach is to compactly represent the high-dimensional wave function $\psixt$ at a single moment in time $t$ using a neural network. The differential equation then indicates how to update in time the full high-dimensional wave function. One would typically optimize a new neural network for time $t+\dd t$ which compactly represents this propagated state.  This is accomplished by optimizing the neural network representation at time $t+\dd t$ to maximally match the non-compact state propagated at first order.  

In this work, we develop an alternative approach.  Instead of building the wave function snapshot by snapshot, we develop an approach which simultaneously compresses the entire space-time wave function.  This is accomplished by having an internal temporal network which takes a time $t$ and outputs the parameters for a second external spatial network which compactly represents the full wave function at that time $t$ as an autoregressive Transformer \cite{luo2021gauge}. The whole spacetime neural network is then optimized using a second order formulation with an implicit mid-point method.  Our work makes a direct connection with the path-integral formulation of quantum mechanics that is used to describe quantum field theories~\cite{Peskin:1995ev}. 

In Sec.~\ref{sec: optimize}, we describe the standard first-order single time-slice approach. We then (Sec.~\ref{sec:s_optimize}~and~\ref{sec:architect}) go on to describe our novel space-time formulation. Our approach (Sec.~\ref{sec:exp}) was tested on a 12-spin Heisenberg model with imaginary time evolution and achieved good performance. Even though the current work is demonstrated with imaginary time quantum dynamics, our approach is general and applies to both real time dynamics as well as more general classes of differential equations.

\section{First Order Optimization and Time-dependent Variational Principle} \label{sec: optimize}

One way to solve the Schrödinger equations in quantum dynamics is to introduce a loss function based on the first order Euler method and minimize the loss function with respect to the neural network representation~\cite{kochkov2018variational}. This can be used as a projective method which iteratively updates the neural network at each time step. For example, in the case of the imaginary time evolution of the Schrödinger equation,
\begin{equation}\label{eq:swo}
    \mathcal{L}_1^{(t)} = \sum_{\mathbf{x}} \abs{\psi(\mathbf{x},t+\dd t) - (I-H \dd t) \psi(\mathbf{x},t)}^2
\end{equation}
where $I$ is the identity operator. For each $t$, $\mathcal{L}_1^{(t)}$ is optimized with respect to $\psi(\mathbf{x},t+\dd t)$ using stochastic gradient descent while $\psi(\mathbf{x},t)$ is fixed without taking gradient. This procedure is then iterated where time $t+\dd t$ becomes the new time $t$. 

Minimizing Eq.~\ref{eq:swo} is equivalent to introducing a nonlinear differential equation in the parameter space~\cite{Yuan_2019}. For a fixed time $t$, by denoting the parameters for the neural network representation of $\psi(\mathbf{x},t)$ as $\theta_t$, the time-dependent variational principle~\cite{Yuan_2019} claims $\frac{d\theta_t}{\dd t} = - F^{-1} \nabla_{\theta_t} E$, 
where $E = \sum_{\mathbf{x}} \psi^{*} (\mathbf{x}, t)H\psixt$ and $F=\sum_{\mathbf{x}} \nabla_{\theta_t} \psi^{*} (\mathbf{x}, t) \nabla_{\theta_t} \psixt$ is the quantum Fisher information matrix. Furthermore, a first order discretization with Euler method gives rise to the gradient update formula with a learning rate of $\dd t$, 
\begin{equation}\label{eq:tdvp}
    \theta_{t+\dd t} = \theta_t - \dd t F^{-1} \nabla_{\theta_t} E
\end{equation}
If one views $E$ as a loss function, the above formula is actually the natural gradient method~\cite{Amari1998} in the context of quantum mechanics. Since the natural gradient method has information beyond first order optimization, it implies that a first order optimization in terms of time step with respect to the function in the wave function space (Eq.~\ref{eq:swo}) gives rises to an optimization beyond first order in the parameter space (Eq.~\ref{eq:tdvp}).

\section{Second Order Optimization with Spacetime Formulation} \label{sec:s_optimize}

An implicit mid-point method which generalizes Eq.~\ref{eq:swo} to second order (Eq.~\ref{eq:midpoint}) has been recently used for real time evolution of the Schrödinger equation~\cite{gutierrez2021real}. Notice that the mid-point method is generic and not limited to the real time Schrödinger equation. It is known that the second order implicit mid-point method has the advantage of preserving the sympletic form of the Hamiltonian dynamics as well as be more stable with larger time step $\dd t$ compared to the first order Euler method~\cite{Hairer2004GeometricNI}. For each $t$, $\mathcal{L}_2^{(t)}$ is optimized with respect to $\psi(\mathbf{x},t+\dd t)$ using stochastic gradient descent while $\psi(\mathbf{x},t)$ is fixed without taking gradient. This procedure is iterative as in the case of the first order optimization with Eq.~\ref{eq:swo}.

\begin{equation}\label{eq:midpoint}
    \mathcal{L}_2^{(t)} = \sum_{\mathbf{x}} \abs{(I+\frac{iH\dd t}{2})\psi(\mathbf{x},t+\dd t) 
    -(I-\frac{iH\dd t}{2})\psixt }^2
\end{equation}

It is a common practice to represent $\psixt$ with a neural network for each $t$ and perform iterative optimization with approaches in Eq.~\ref{eq:swo},~\ref{eq:tdvp},~\ref{eq:midpoint} in the community of neural network quantum states~\cite{Carleo_2017,kochkov2018variational,gutierrez2021real}. The neural network in the above approaches essentially only represents the spatial part of the wave function because different neural networks are required for different $t$. Although the above approaches are flexible and useful, the number of copies of neural network increases linearly with the number of time step. 

\begin{figure*}[h!]
\centering
\includegraphics[width=0.85\linewidth]{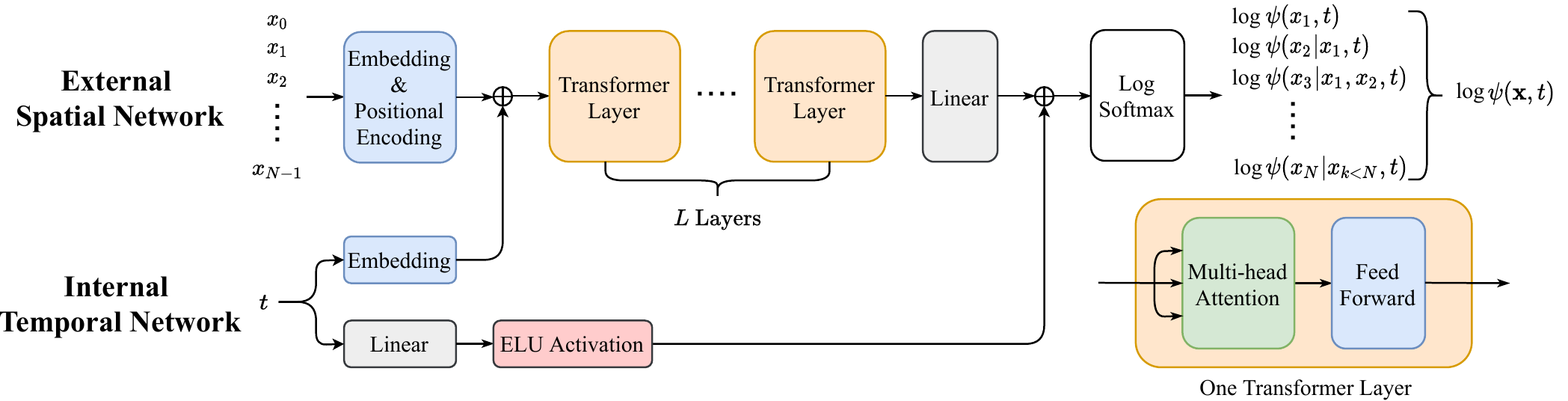}
\caption{Architecture of autoregressive spacetime neural network. The neural network has an external spatial network and an internal temporal network. The internal temporal network takes the time as the input and generates the relevant weights for the external spatial network. The external spatial network is a Transformer which utilizes these weights to generate the log wave function.}
\label{fig:model_architecture}
\end{figure*}

In this work, we propose a second order optimization method with spacetime formulation, which makes use of the implicit mid-point method and the full spatial and temporal neural network representation.  Instead of parameterizing the wave function $\psixt$ with a neural network for each discrete time point $t$, we construct a spacetime neural network that represents $\psixt$ for all $\mathbf{x}$ and $t$ altogether. We further utilize the second order implicit mid-point formulation and define a new loss. For example, in the context of imaginary time evolution of the Schrödinger equation over a total time $T$, we have 
\begin{equation}\label{eq:spacetime}
    \mathcal{L}_T = \sum_{t}^{T} \mathcal{L}_t((I+\frac{H\dd t}{2})\psi(\mathbf{x},t+\dd t), (I-\frac{H\dd t}{2})\psixt)
\end{equation}
where $\mathcal{L}_t(\psi_{1}(\mathbf{x},t), \psi_{2}(\mathbf{x},t))$ is a loss function between two wave functions $\psi_{1}(\mathbf{x},t)$ and $\psi_{2}(\mathbf{x},t)$ for a fixed $t$. In this work, we choose $\mathcal{L}_t$ to be the log overlap function such that 
\begin{equation}
    \mathcal{L}_t(\psi_{1}(\mathbf{x},t), \psi_{2}(\mathbf{x},t)) = -\log \frac{|\langle \psi_1, \psi_2\rangle|^2}{\langle \psi_1,\psi_1\rangle \langle\psi_2,\psi_2\rangle}
\end{equation}
where $\langle \psi_{1}, \psi_{2} \rangle =\sum_{\mathbf{x}} \psi_{1}^{*}(\mathbf{x}, t) \psi_{2}(\mathbf{x}, t)$.
Notice that the key difference between Eq.~\ref{eq:spacetime} and Eq.~\ref{eq:midpoint} is that Eq.~\ref{eq:spacetime} is defined over all time $t$ and $\psixt$ is represented by only one spacetime neural network. For $T=\dd t$ as one time step, Eq.~\ref{eq:spacetime} will reduce to the imaginary time version of Eq.~\ref{eq:midpoint} by choosing $\mathcal{L}_t$ as the $L_2$ norm. Fig.~\ref{fig:spacetime} demonstrates the high level picture of the second order optimization spacetime neural network. Our spacetime neural network consists of internal temporal network and external spatial network and the details will be given in the following section. The internal temporal network generates a series of external spatial networks to represent a set of $\{\psi(\mathbf{x}, t_i)\}$, which are used in the loss function Eq.~\ref{eq:spacetime}. To optimize Eq.~\ref{eq:spacetime} with high dimensional $\psixt$, we use exact sampling for configuration $\mathbf{x}$ and take a uniform discretization in $[0,T]$ with step $\dd t$ and sum over $t$. 

The formulation of Eq.~\ref{eq:spacetime} provides a second order method for optimizing the spacetime neural network $\psixt$, which is able to predict values of $\psixt$ for any $\mathbf{x}$ and $t$ after the optimization. Even though the main discussion of our work is in the context of high dimensional 
Schrödinger equation, the approach could be extended to other high dimensional differential equations.

\section{Architecture of Spacetime Neural Network} 
\label{sec:architect}

In this section, we propose a self-normalized spacetime neural network architecture for solving the high dimensional Schrödinger equation. Consider a wave function $\psixt$ of $N$ spin particles for a fixed $t$; its dimensionality grows exponentially as $2^N$. In order to resolve the high-dimensionality issue presented by quantum mechanics, we need a compact representation for the wave function. In addition, the wave function is a complex-valued function that is $L_2$ normalized, which requires that $\sum_{\mathbf{x}} |\psixt|^2 = 1$ $\forall t$.
For a wave function $\psi(\mathbf{x})$ with only spatial dependence, recent progress~\cite{Sharir_2020} in autoregressive models provides a method for representing $\psi(\mathbf{x})$ with polynomial number of parameters and maintaining the normalization condition $\sum_{\mathbf{x} }|\psi(\mathbf{x})|^2=1$. The key idea behind the autoregressive model is to factorize the 
wave function into conditional wave functions on previous sites, such that $\psi(\mathbf{x})=\psi(x_1,...,x_N) = \prod_{k=1}^{N} \psi(x_k|x_{k-1},...,x_1)$.
One can normalize the high dimensional wave function $\psi(\mathbf{x})$ by normalizing all the conditional wave function distributions $\sum_{x_1,...,x_k} |\psi(x_k|x_{k-1},...,x_1)|^2 = 1$ for all $k \leq N$. In this work, we generalize the autoregressive model to a spacetime neural network wave function such that
\begin{equation}\label{eq:network}
    \psi(\mathbf{x},t)=\prod_{k=1}^{N} \psi(x_k|x_{k-1},...,x_1,t)
\end{equation}
This construction ensures $\psixt$ is normalized for each $t$. Therefore, for each $t$, $\mathbf{x}$ can be sampled exactly according to the conditional probability distribution of the wave function, making it more efficient compared to Markov chain Monte Carlo sampling for high dimensional wave functions.

To realize Eq.~\ref{eq:network}, we construct the spacetime neural network with an internal temporal network and an external spatial network as Fig.~\ref{fig:model_architecture} shows. For the external spatial network, we use a Transformer since it has the desired autoregressive property and can compactly represent high dimensional distributions. Transformers \cite{transformer} were introduced in 2017 for natural language processing and later were found applicable to many different areas. The external spatial network consists of an embedding layer, a positional encoding layer, $L$ transformer layers ($L=1$ in our experiment), a linear layer and a log softmax layer. Each transformer layer is composed of a multi-head attention layer and a feed-forward layer. For our implementation, we choose $16$ as the hidden dimension for all transformer layers and linear layers, and we set the number of attention heads to $8$. The internal temporal network takes the current time $t$ as input and adds time dependencies to the spatial network after the embedding layer and before the log softmax layer. The neural network takes the $N$-particle configuration $\mathbf{x} = (x_0, x_1,...,x_{N-1})$ and the current time $t$ as inputs, and outputs the log of the wave function $\psi(\mathbf{x},t)$ at time $t$.

\section{Numerical Experiments} \label{sec:exp}

We benchmarked our spacetime neural network on a Heisenberg model with $12$ spins for imaginary time evolution. The Hamiltonian $H$ of the Heisenberg model is given by:
\begin{equation}
    H = -J\left(\sum_{i}\sigma^x_i\sigma^x_{i+1} + \sum_{i}\sigma^y_i\sigma^y_{i+1} \right) + \sum_{i}\sigma^z_i\sigma^z_{i+1}
\end{equation}
where $J$ is the coupling constant and $\sigma_i^x, \sigma_i^y, \sigma_i^z$ are Pauli matrices acting on site $i$. We set $J=1$, $\dd t=0.01$ and $T = 2$ for solving $\psixt$, where $\mathbf{x}$ is the spin configuration of the 12 sites. The optimization is performed using Adam optimizer for 150 steps.

\begin{figure}[h!]
\includegraphics[width=\linewidth]{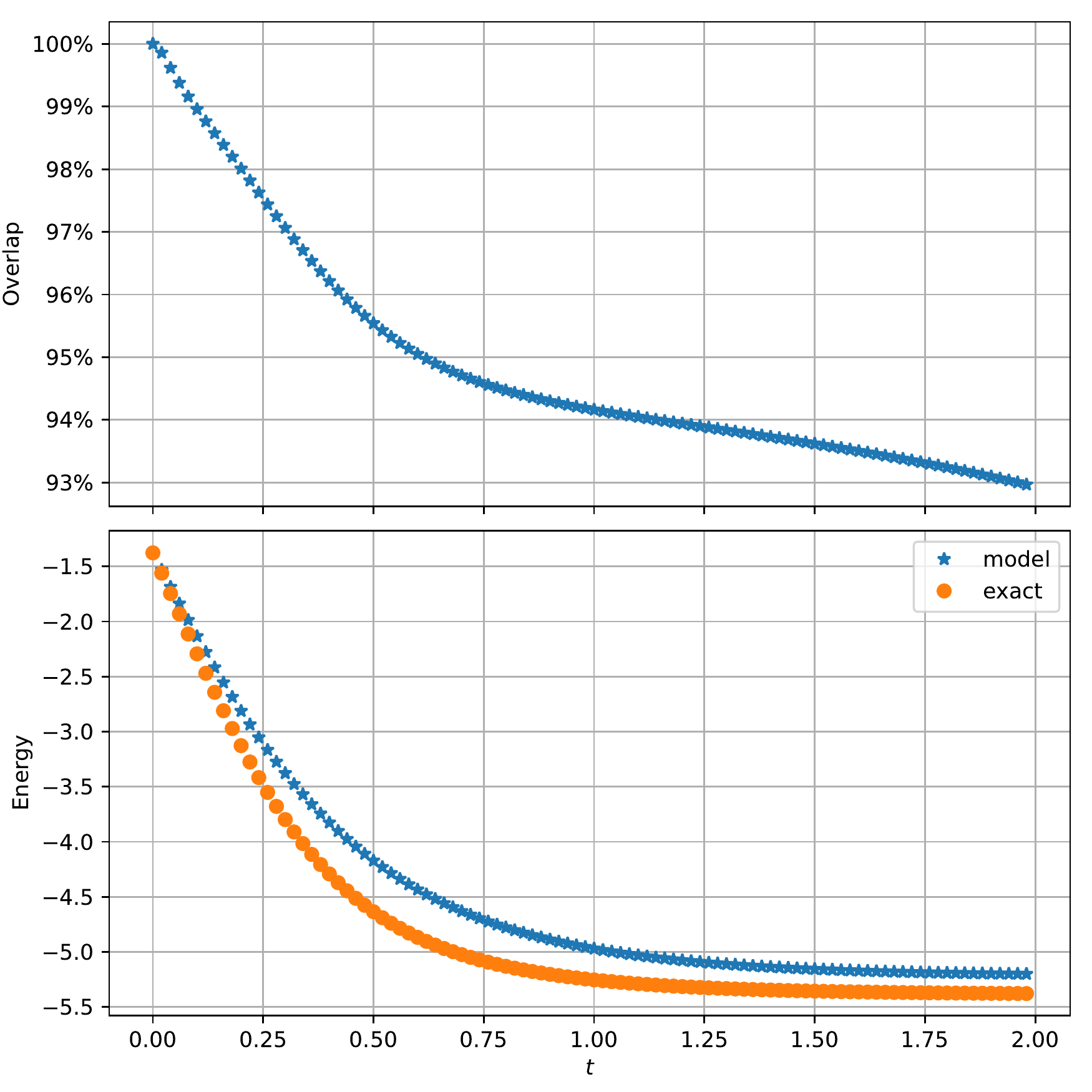}
\caption{Overlap and energy comparison between spacetime neural network and the exact solution on a 12 spins Heisenberg model.}
\label{fig:results}
\end{figure}

We use the absolute overlap value between the spacetime neural network and the exact solution, i.e. $|\sum_{\mathbf{x}} \psi^{*} (\mathbf{x}, t) \psi_{\mathrm{exact}}(\mathbf{x}, t)|$, to measure the accuracy of the simulation of imaginary time evolution. The exact solution is obtained from exact diagonalization. We also compute the exact and the predicted energy for the wave functions at different time steps. The energy is calculated as $E(t) = \sum_{\mathbf{x}} \psi^{*} (\mathbf{x}, t)H\psixt$, 
where $H$ is the Hamiltonian of the system. Our experiment results are shown in Fig. \ref{fig:results} where the top plot shows the overlap and the bottom plot shows the energy. It is found that our method provides good overlap and close match in the energy to the exact solution.

\section{Discussions and Conclusion}

We have introduced a spacetime neural network with a second order optimization for solving high dimensional Schrödinger equations. The advantages of the approach come from optimizing $\psixt$ for all $\mathbf{x}$ and $t$ with a second order optimization formulation. Our spacetime neural network is autoregressive and self-normalized for each $t$, which enables efficient exact sampling in high dimension. Contrasted to the standard method of obtaining the wave function at discrete times with iterative projective method, our spacetime neural network $\psixt$ stores all spatial and temporal information, which is indeed equivalent to obtaining the path integral in quantum mechanics. It further allows to compute observables correlated in time, such as the Green function.

Even though the current experiments are performed for quantum dynamics with the imaginary time Schrödinger equation, our approach is general and it could be applied to differential equations in both quantum and classical physics contexts. One next step is to explore the formulation in the situation of real time quantum dynamics, where there could be more oscillation in the dynamics. Our work is also connected to recent work~\cite{RAISSI2019686,Sirignano_2018,Han8505} in the applied mathematics community which aims to solve differential equations $\frac{\partial u(\mathbf{x},t)}{\partial t} = Lu(\mathbf{x},t)$ ($L$ is an operator) by parameterizing $u(\mathbf{x},t)$ with neural networks. It will be interesting to apply our approach to different scenarios, such as the Hamilton-Jacobi equation, the high-dimensional master equation and the  Black-Scholes model. It is also anticipated that improvements on the architecture of the spacetime neural network will be helpful for solving differential equations in different setups. 

Another interesting direction is to explore the connection between our spacetime neural network and the neural ODEs~\cite{chen2019neural}. One can use neural ODEs to generate the spacetime neural network for different $t$ and consider the optimization using adjoint method. We believe that our work opens up new opportunities for research in machine learning, applied mathematics, and physics. 

\section*{Acknowledgements}
This work utilizes resources supported by NSF through the Major Research Instrumentation program OAC-1725729 as well as the University of Illinois at Urbana-Champaign~\cite{hal}. ZZ is partially supported by NSF OAC-1934757 and Alfred P. Sloan Foundation. VMH is partially supported by NSF DMS-1452597 and DMS-2009981. BKC acknowledges
support from the Department of Energy grant DOE DESC0020165.

\bibliography{references}
\bibliographystyle{icml2021}

\end{document}